\documentclass[aps,prb,twocolumn,superscriptaddress,msmath,amssymb]{revtex4}

\usepackage{graphicx,bm}
\begin{document}

\title{Fincher-Burke spin excitations and $\omega/T$ scaling in insulating La$_{1.95}$Sr$_{0.05}$CuO$_4$}

\author{Wei Bao}
\affiliation{Condensed Matter and Thermal Physics, Los Alamos National Laboratory, Los Alamos, NM 87545}
\author{Y. Chen}
\affiliation{NIST Center for Neutron Research, National Institute of Standards 
and Technology, Gaithersburg, MD 20899}
\affiliation{Dept.\ of Materials Science and Engineering, University of
Maryland, College Park, MD 20742}
\author{K. Yamada}
\affiliation{Institute of Materials Research, Tohoku University, Sendai 980-8577, Japan}
\author{A. T. Savici}
\affiliation{Brookhaven National Laboratory, Upton, NY 11973}
\author{P. L. Russo}
\affiliation{TRIUMF, Vancouver, BC V6T 2A3, Canada}
\author{J. E. Lorenzo}
\affiliation{Institut N\'{e}el, CNRS, BP 166X, F-38043, Grenoble, France}
\author{J.-H. Chung}
\affiliation{NIST Center for Neutron Research, National Institute of Standards 
and Technology, Gaithersburg, MD 20899}
\affiliation{Dept.\ of Materials Science and Engineering, University of
Maryland, College Park, MD 20742}

\date{\today}

\begin{abstract}
Insulating La$_{1.95}$Sr$_{0.05}$CuO$_4$ shares with superconducting cuprates the same
Fincher-Burke spin excitations, which usually are observed in itinerant antiferromagnets. 
The local spectral function satisfies $\omega/T$ scaling above $\sim$16 K for this incommensurate insulating cuprate. Together with previous results in commensurate insulating and incommensurate superconducting cuprates, these results further support the general scaling prediction for square-lattice quantum spin $S=1/2$ systems. The width of incommensurate peaks in La$_{1.95}$Sr$_{0.05}$CuO$_4$ scales to a similar finite value as at optimal doping, strongly suggesting that they are similarly distant from a quantum critical point. They might both be limited to a finite correlation length by partial spin-glass freezing.

\end{abstract}


\maketitle

It is well known that the ($\pi,\pi$) peak of the N\'{e}el antiferromagnetic order of the parent compounds is replaced by a quartet of incommensurate peaks when cuprates are sufficiently doped to have become superconducting\cite{john_review}. The very recent excitement is that spin excitations measured in
superconducting La$_{2-x}$Sr$_x$CuO$_4$ (LSCO) ($x=0.10$ and
0.16)\cite{lsco_nbc}, YBa$_2$Cu$_3$O$_{6.6}$
(YBCO)\cite{ybco_smh}, as well as in the ``stripe-ordered''
La$_{1.875}$Ba$_{0.125}$CuO$_4$ with a suppressed $T_C$\cite{John_LaBa}
share a common spectral feature: broad excitation continua, originating from the incommensurate quartet, disperse towards the ($\pi,\pi$) point with increasing energy at the rate of the spin-wave velocity of the parent compounds. The spectrum is distinct from the spin-waves in two important ways: 1) the excitations are not resolution-limited; 2) there are no the outward branches. Such a spin excitation spectrum previously has been observed in itinerant spin-density-wave antiferromagnets such as elemental metal Cr\cite{FincherMode} and strongly correlated metal V$_{2-y}$O$_3$\cite{bao96b}, and the single-lobed dispersive continuum is referred as the Fincher-Burke mode. 
A self-consistent theory has been developed by Moriya and others to describe the mode\cite{scrmorb1}, and quantitative agreement has been achieved for three-dimensional itinerant antiferromagnets\cite{bao96b}. Itinerant theories have also been developed to
account for the Fincher-Burke-like modes in superconducting cuprates\cite{rpa_ara}.

Sandwiched between the parent antiferromagnetic insulator and the high-$T_C$ superconductor in the phase diagram of LSCO is a distinct doping regime from $x=0.02$ to 0.055\cite{Waki_Sr3}. Cuprates in the regime are insulators without the long-range N\'{e}el order, and a spin-glass transition occurs at $T_f \lesssim 10$~K\cite{bjbquasi,fchou}. 
This doping range is often referred to as the spin-glass
regime, although the spin-glass phase extends to
both lower and higher dopings in the N\'{e}el and superconducting
states.  Magnetic correlations were extensively investigated and regarded as being {\em
commensurate}, as in the parent compound\cite{la2bkb,la2mm}.  However, with
improved single-crystal samples, magnetic correlations show a
novel {\it incommensurate} doublet, which also differs distinctly from the quartet in the superconductors\cite{Waki_Sr3,L214_sw}. In this paper, we
report that spin excitations of La$_{1.95}$Sr$_{0.05}$CuO$_4$ in the {\em insulating} spin-glass regime are also composed of the Fincher-Burke modes, originating from the incommensurate doublet, with a velocity the same as in superconducting LSCO.
Although cuprates in the spin-glass regime are insulators, they are not 
the usual band insulators, and part of the Fermi surface
may have survived\cite{dmft_MLT,arpes_lsco}. It would be interesting to investigate
whether the Fincher-Burke modes reported here can be accounted for by extending 
theories for similar spin excitations of superconducting cuprates.

Meanwhile, for quasi-two-dimensional (2D) spin $S$=1/2 cuprates, the temperature range where spin fluctuations are investigated is within $T<<J/k_B\sim 1000$~K\cite{la2smhao}, where classical statistical mechanics has to be replaced by quantum statistical mechanics\cite{qpt_inv}. The general scaling argument of 2D quantum  statistical systems leads to a prediction for samples which are not exactly at a quantum critical point (QCP) that for $T\ll J/k_B$ but above a low-temperature limit $T_X$,  the energy scale for 2D spin fluctuations at long wavelengths is $k_B T$\cite{qpt_inv,2dheiqd}.  Hence, for $T_X<T\ll J/k_B$, the spin excitation spectrum follows the $\omega/T$ scaling\cite{2dheiqd}, which is not a robust feature of the Moriya theory\cite{scrmorb1}.  For $T<T_X$, a constant energy gap is predicted\cite{2dheiqd}. We test
these predictions against the distinct incommensurate spin excitations
from the doublet in La$_{1.95}$Sr$_{0.05}$CuO$_4$.  We also compare our results with previous investigations of incommensurate spin fluctuations from the quartet in superconducting cuprate\cite{la2gas} and commensurate ones
at the ($\pi,\pi$) point in insulating cuprate\cite{bao02c}.  Our data support the
scaling above $T_X$ but no gap is observed below $T_X$.  Surprisingly, scaling analysis of the $\bm{q}$ and $\omega$ dependent spin
excitation spectra indicates that La$_{1.95}$Sr$_{0.05}$CuO$_4$ and optimally doped LSCO are similarly
distant from the QCP.

\begin{figure}[tb]
\includegraphics[width=.75\columnwidth,angle=90,clip=true]{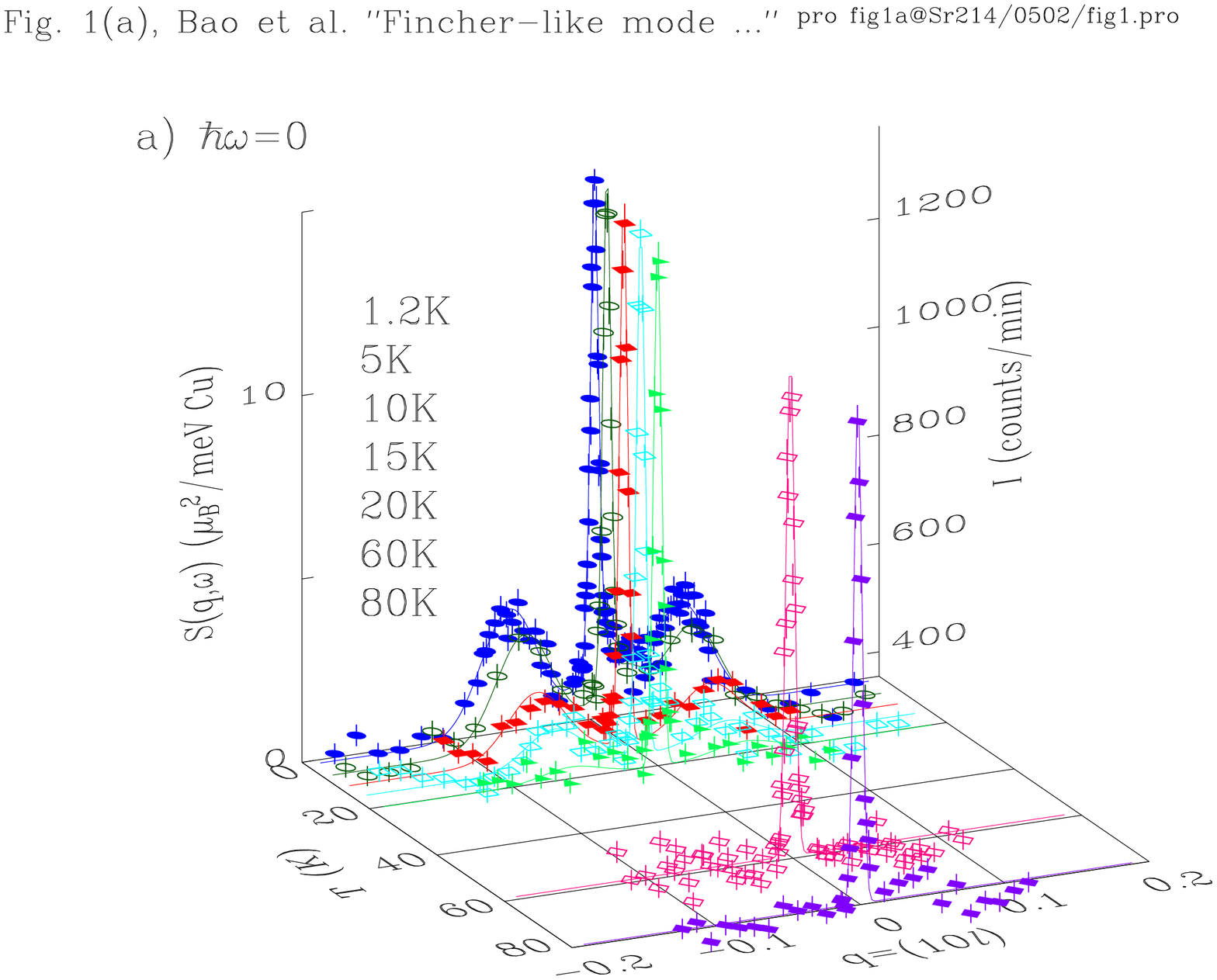}
\vskip -0.3 cm
\includegraphics[width=.452\columnwidth,angle=90,clip=true,trim=0 0 0 0]{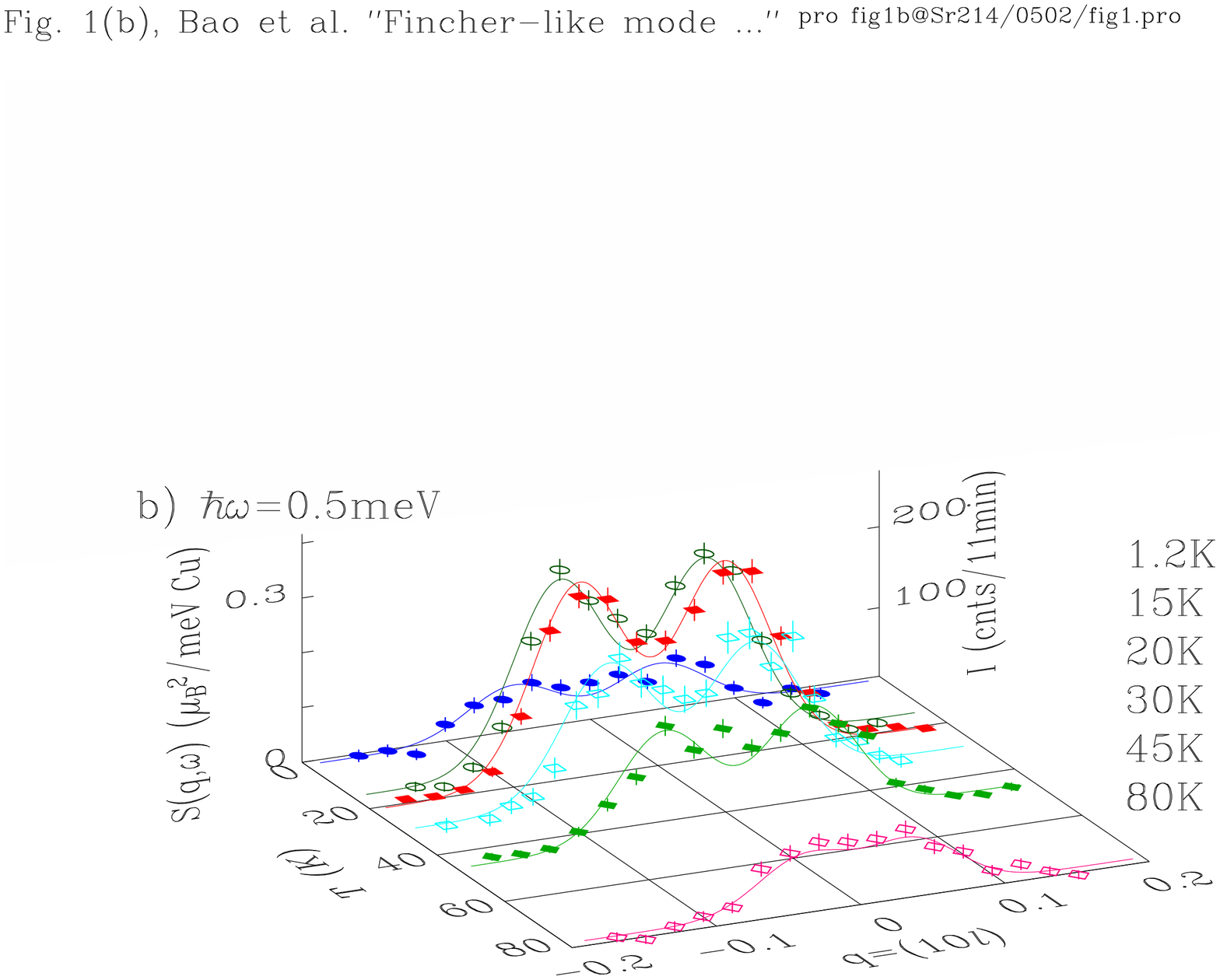}
\vskip -2 cm
\includegraphics[width=.68\columnwidth,angle=90,clip=true]{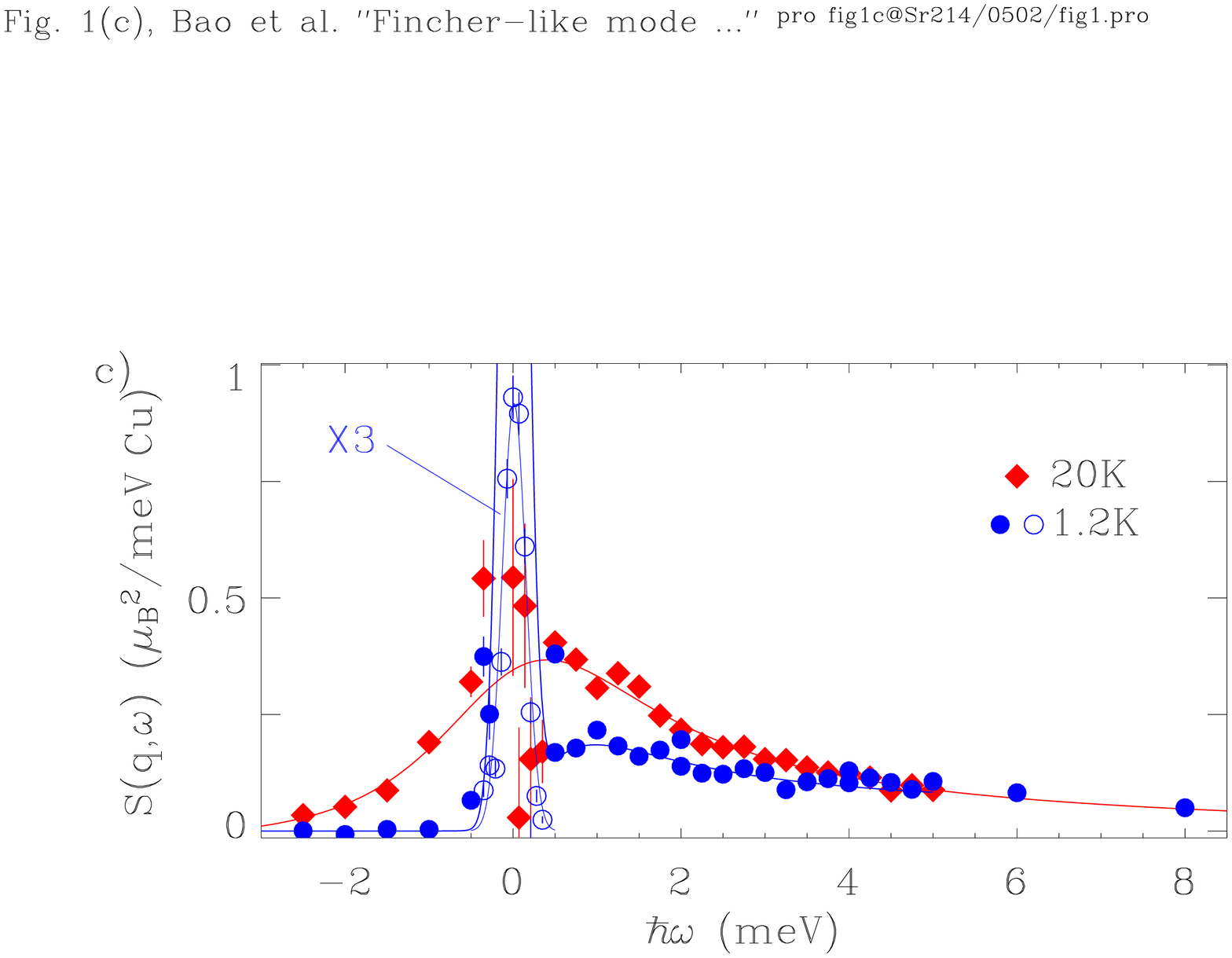}
\vskip -.5 cm
\caption{\label{fig1} (color online)
The generalized spin correlation function $S(\bm{q},\omega)$ as a function
of $\bm{q}$ along the $c$-axis measured at (a) $\hbar \omega=0$ and
(b) 0.5 meV, respectively, at various temperatures. 
(c) $S(\bm{q},\omega)$ as a function
of energy at $\bm{q}=(1,0,0.05)$ and at 1.2 and 20 K, respectively.
The ``central peak'' at 1.2 K is also shown on a 1/3 scale with open circles. 
Background has been 
subtracted in (c).
}
\end{figure}

A single piece of La$_{1.95}$Sr$_{0.05}$CuO$_4$ crystal of 5.2 g was used in this work. It was grown using a traveling-solvent floating-zone
method as described previously\cite{Waki_Sr3,L214_sw}. We use the 
orthorhombic $Cmca$ unit cell ($a=5.338\AA$, $b=13.16\AA$,
$c=5.404\AA$ at 1.5~K) to describe the $\bm{q}$-space for measurements
at NIST using the cold neutron triple-axis spectrometer SPINS. 
The sample temperature was controlled by a pumped He$^4$ cryostat.
The horizontal collimations before and after the sample
were both 80$^{\prime}$, and a Be filter cooled by liquid nitrogen
was used after the sample to reduce higher order neutrons
passing through the pyrolytic
graphite (002) used for both monochromator and analyzer.
The intensity of magnetic neutron scattering was counted against 
a flux monitor placed before the sample in a 
fixed $E_f=5$~meV configuration and normalized to yield $S(\bm{q},\omega)$
in absolute units.
Such a cold neutron spectrometer readily resolves the
incommensurate doublet near (100), i.e., the ($\pi,\pi$) point (see Fig.~\ref{fig1}),
while it is difficult to resolve the doublet using a thermal 
neutron triple-axis spectrometer due to its coarser resolution. Hence,
we will present only the cold neutron scattering results here.

We first present the {\em nominal} elastic signal at various temperatures.
Fig.~\ref{fig1}(a) shows constant-energy $\hbar \omega$=0 scans through the incommensurate
doublet from 1.2 to 80 K.
The sharp peak at (100) is due to higher-order diffraction of (200).  
Its width indicates the instrument
resolution.  The magnetic doublets at $\bm{q}_{IC}$=(1,0,$\pm 0.058(2)$) are
obviously broader than the resolution.  The deconvoluted peak width 
yields the in-plane correlation length for the nominally elastic spin correlations, $\xi_{\Box}=34(2) \AA$ at 1.2 K, about 9 nearest-neighbor Cu spacings.
With increasing temperature, the doublet monotonically decreases in intensity without
appreciable change in either the peak width or position.  Above $\sim$ 20 K, the doublet disappears, consistent with previous 
studies\cite{Waki_Sr3}. 

At finite energies, however, the temperature dependence of the doublet is entirely different from  that at $\hbar \omega=0$. Fig.~\ref{fig1}(b) shows
constant-energy $\hbar \omega$=0.5 meV scans measured in the same temperature range.
Instead of monotonically decreasing, the intensity first increases,
reaches a maximum between 15 and 20 K, and then decreases with further rising temperature.
The intensity at the peak shoulder, e.g., at $\bm{q}=(1,0,0.2)$, in Fig.~\ref{fig1}(a)-(b) measures a temperature-independent background, which has been subtracted
in Fig.~\ref{fig1}(c).
Note that at 1.2 K, $S(\bm{q},\omega)$ at 0.5 meV is more than one
order of magnitude weaker than at $\hbar \omega=0$. This reflects the fact
that the energy spectrum of $S(\bm{q},\omega)$ shows a prominent sharp ``central peak'' at
$\omega=0$ at low temperatures, see Fig.~\ref{fig1}(c). 
The ``central peak'' is energy-resolution-limited at the SPINS spectrometer, with the full-width-at-half-maximum (FWHM) of 0.3 meV.
However, the nominal elastic signal from La$_{1.95}$Sr$_{0.05}$CuO$_4$
is not truly static at $T>T_g\approx 5$ K as determined by our $\mu$SR measurements, 
which has an energy resolution of $\sim$10$^{-6}$ meV.
Details of the $\mu$SR study will be published elsewhere.
Similar ``central peak'' phenomenon has been reported 
for Li-doped La$_2$CuO$_4$ and YBa$_2$Cu$_3$O$_{6+x}$, and the 
very slow spin dynamics is attributed to
a partial spin-glass transition\cite{bao05a,stock}.

\begin{figure}[t]
\includegraphics[width=.73\columnwidth,angle=90,clip=true]{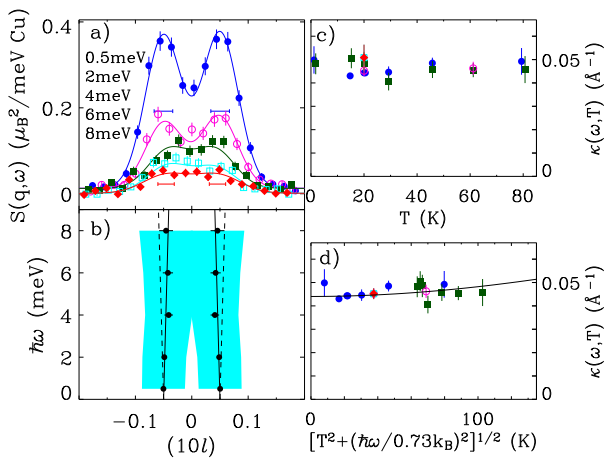}
\vskip -.5cm
\caption{\label{fig2} (color online)
(a) Constant-$\hbar\omega$ scans at 20 K. The solid lines are two equal-width gaussians. The horizontal bars indicate instrument resolution (FWHM). 
(b) The double peak positions. The shaded area
represents the FWHM. The spectral shape strongly resembles that for 
La$_{1.84}$Sr$_{0.16}$CuO$_4$\cite{lsco_nbc} and V$_{1.97}$O$_3$\cite{bao96b}.
The lines denote the spin-wave branches with a velocity of 850 
meV$\AA$ of La$_2$CuO$_4$\cite{la2smhao}.
The FWHM/2 measured at those energies listed in (a) is shown as a function of
temperature in (c) and reduced temperature in (d).
}
\end{figure}

What is the energy dependence of the doublet?
Fig.~\ref{fig2}(a) shows scans at various energies at 20 K.
Below 3 meV, the two incommensurate peaks are clearly distinguishable.
As energy increases, the doublet merges into a flat-top peak. 
The scans in Fig.~\ref{fig2}(a) can be fitted using two gaussians of the same width. The peak positions are shown as the black circles in Fig.~\ref{fig2}(b).
The dispersion is consistent with the inner branches of spin-waves (solid lines)\cite{la2smhao}. The same dispersion rate has been reported for
superconducting La$_{2-x}$Sr$_x$CuO$_4$ ($x=0.10$ and 0.16)\cite{lsco_nbc}.
The shaded area in Fig.~\ref{fig2}(b) covers the FWHM, which grows slowly with energy
from 0.089(3) $\AA^{-1}$ at 0.5 meV. The width is comparable to that for the
$x=0.16$ sample. Because of the smaller doublet separation, the
merge of the peaks occurs at $\sim$4 meV, much lower than at $\sim$40 meV
for La$_{1.84}$Sr$_{0.16}$CuO$_4$\cite{lsco_nbc}. 
Thus, despite very different ground states and spatial spin correlations, insulating and superconducting LSCO share the common Fincher-Burke spectral shape.

\begin{figure}[tb]
\includegraphics[width=.7\columnwidth,angle=90,clip=true]{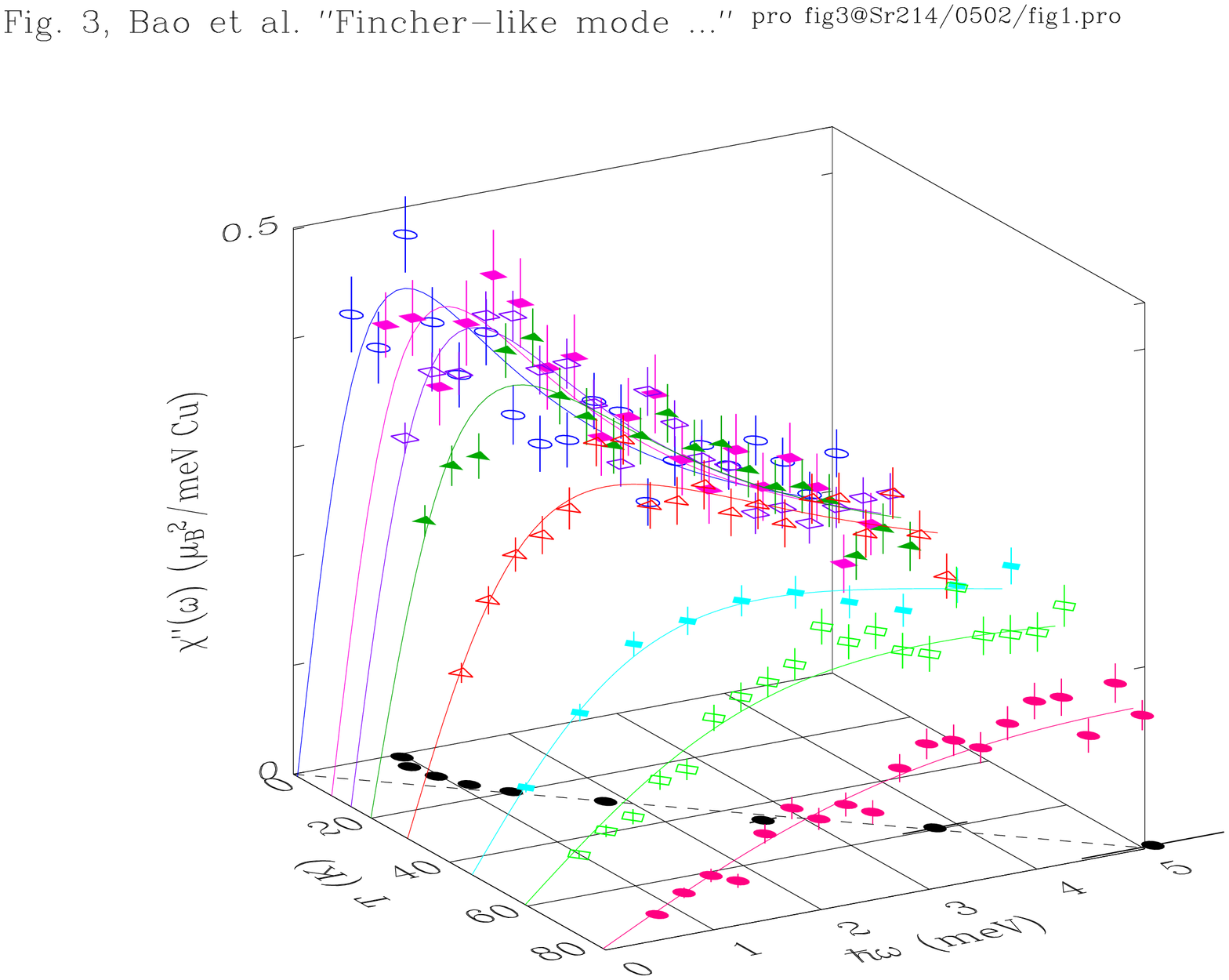}
\vskip -8mm
\caption{\label{fig3} (color online)
The local dynamic magnetic susceptibility $\chi''(\omega)$ measured at various temperatures.
The solid lines are the least-square fit to Eq.~(\ref{debye}).
The relaxation constant $\Gamma$ is shown on the $\hbar\omega$-$T$ plane
as a function of $T$, and departs from the dashed line $\Gamma= 0.73 k_B T$ below
$T_X\approx 16$ K.
}
\end{figure}

Now we turn to examination of scaling behavior of spin excitations. 
Historically, it was done in the spin-glass regime through the local spin correlation function
$S(\omega)=\int d\bm{q} S(\bm{q},\omega)$ using
cuprate samples showing {\em commensurate} spin correlations\cite{la2bkb,la2mm}.
It is re-examined here with the very different {\em incommensurate} spin 
correlations of our improved sample.  
The imaginary part of the local dynamic magnetic susceptibility
relates to $S(\omega)$ by the fluctuation-dissipation theorem
\begin{equation}
\chi''(\omega)=\pi (1-e^{-\hbar\omega/k_BT}) S(\omega).
\label{fdt}
\end{equation}
Fig.~\ref{fig3} shows $\chi''(\omega)$ from 1.2 to 80 K,
which is well described by the Debye relaxor model
\begin{equation}
\chi''(\omega)=\frac{\chi_0 (\hbar \omega/\Gamma )}{1+(\hbar \omega/\Gamma)^2},
\label{debye}
\end{equation}
where $\chi_0$ is the local staggered static magnetic susceptibility, and
$\Gamma$ the spin relaxation constant. 
In previous studies\cite{la2bkb,la2mm},
the local $\chi''(\omega)$ was modeled by 
\begin{equation}
\chi''(\omega)=I(\omega)\arctan[a_1(\hbar \omega/k_B T )+a_2 (\hbar \omega/k_B T )^2].
\end{equation}
The $\arctan$ function is stipulated by the marginal Fermi liquid 
theory\cite{la2bkb,bibmfl},
but $I(\omega)$ has no determined analytic form\cite{la2mm}. Hence, we opt for the well-known
Debye relaxor model, Eq.~(\ref{debye}), to fit our data.  The Debye relaxor model has also successfully described measured $\chi''(\omega)$ of insulating La$_{2}$Cu$_{0.94}$Li$_{0.06}$O$_4$, which has {\em commensurate} magnetic correlations\cite{bao02c}.

\begin{figure}[tbh]
\includegraphics[width=.6\columnwidth,angle=90,clip=true]{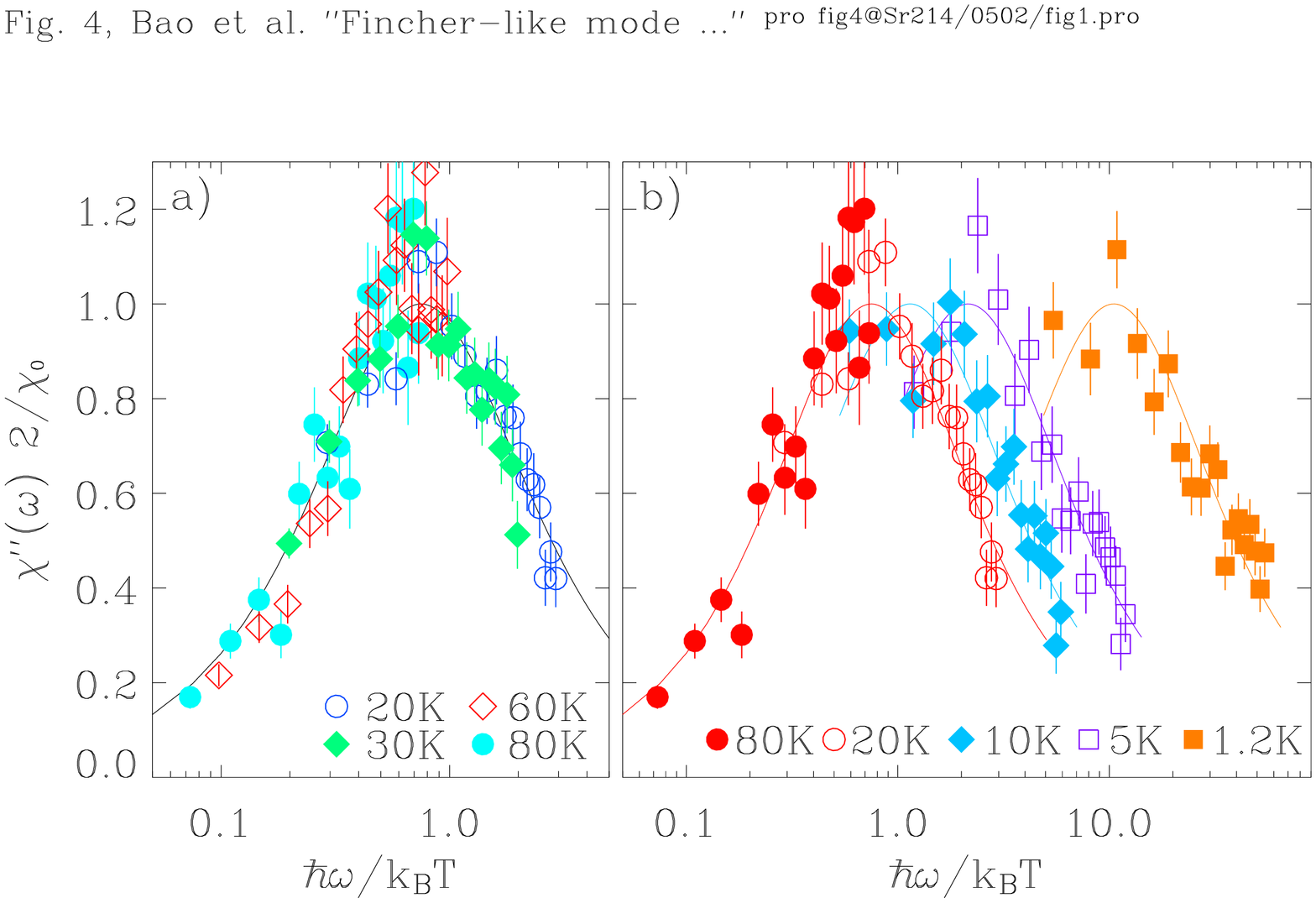}
\vskip -1.2 cm
\caption{\label{fig4} (color online)
The normalized local dynamic magnetic susceptibility $\chi''(\omega) 2/\chi_0$ as a
function of $\hbar\omega/k_BT$. (a) The $\omega/T$ scaling is followed for
data taken above $T_X\simeq 16$~K. 
The solid curve is the scaling function $y=2/[1+(x/0.73)^2]$. 
(b) Data taken below $T_X$ does not follow the scaling function.}
\end{figure}

On the base plane of Fig.~\ref{fig3}, the spin relaxation constant $\Gamma$ obtained
from the least-squares fit is shown as a function of temperature.
The good instrument resolution,
0.3 meV (FWHM), has a negligible effect during fitting.  
One interesting result is that $\Gamma= 0.73(2) k_B T$ for temperatures above $T_X\approx 16$ K.
Hence, when $\chi''(\omega)$, normalized by
its maximum $\chi_0/2$ at $\hbar\omega=\Gamma$, is plotted as a function of 
$\hbar\omega/k_B T$,  Eq.~(\ref{debye}) dictates that all data collected above $T_X$ 
collapse onto a single universal curve $y=2/[1+(x/0.73)^2]$ and Fig.~\ref{fig4}(a)
bears this out. The result is commonly referred to as the $\omega/T$ scaling, and $\Gamma/k_B T = O(1)$ is a hallmark of quantum magnetic 
theory\cite{qpt_inv,2dheiqd}.  For $T<T_X$, Fig.~\ref{fig3} shows that $\Gamma$ 
departs from the proportionality to temperature.  Consequently, the low 
temperature data would not follow the scaling curve, as demonstrated by Fig.~\ref{fig4}(b).  Note that the spectral function Eq.~(\ref{debye}) does not become gapped below $T_X$, contrary to non-random quantum theory\cite{2dheiqd}, but can be explained by dopant scattering\cite{2dheiz2,2dheisu}.

The $\omega/T$ scaling and its departure below $T_X$ shown in 
Fig.~\ref{fig4} for La$_{1.95}$Sr$_{0.05}$CuO$_4$ bears a striking similarity 
to what reported for La$_{2}$Cu$_{0.94}$Li$_{0.06}$O$_4$\cite{bao02c}.
The two cuprates have similar hole concentration and develop spin-glass
at similar $T_g$. However, they differ in several important ways:
i) The dopants are out of the CuO$_2$
plane in the Sr compound, but directly replace Cu$^{2+}$ in the Li compound.
ii) The former becomes a superconductor with additional 0.5\% more holes, 
but the latter always remains an insulator. 
iii) Magnetic correlations are incommensurate in the former,
but commensurate in the latter. 
iv) The in-plane correlation length $\xi_{\Box}\simeq 34(2) \AA$ for the glassy
spin component in the former,
but $\xi_{\Box}\gg 274\AA$ in the latter, and
the $\kappa(\omega, T)$ shown 
in Fig.~\ref{fig2}(c) is more
than double that in the latter\cite{bao05a}.
v) $\Gamma/k_B T= 0.73$ for the former, and 0.18 for the latter\cite{bao02c}.
In spite of these differences, $\Gamma$ saturates at $\Gamma_0\sim 1$ meV 
and $\chi''(\omega)$ becomes essentially 
$T$-independent (see Fig.~\ref{fig3}) for both cuprates below $T_X$.
As a consequence, Eq.~(1) requires a reduced $S(\omega)$ at
low energies when the temperature decreases  below $T_X$, as observed in Fig.~\ref{fig1}(b) and (c), 
in sharp contrast to a magnet at the QCP.

Scaling of spin excitations has also been examined near the optimal doping
for La$_{1.86}$Sr$_{0.14}$CuO$_4$\cite{la2gas}. The material is concluded to be near a QCP, namely, $T_X$$\rightarrow$0, with some caveat\cite{la2sc}.
The $\chi''_P/\omega$ in [\cite{la2gas}], equaling to $\chi_0/\Gamma$ of Eq.~(\ref{debye}), would saturate below $T_X$,
but $T_C=35$ K sets the upper limit for measurable
$T_X$ in La$_{1.86}$Sr$_{0.14}$CuO$_4$. Hence it cannot be determined
whether La$_{1.86}$Sr$_{0.14}$CuO$_4$ or La$_{1.95}$Sr$_{0.05}$CuO$_4$
is closer to a QCP with a lower $T_X$. Another method to assess
the distance from the QCP is to examine the width 
of constant-$\hbar\omega$ scans, see Fig.~\ref{fig2}(c). 
Adapting the ansatz in [\cite{la2gas}], the $\kappa(\omega, T)$ plotted
as a function of a reduced temperature better collapses our data in Fig.~\ref{fig2}(d), 
and the solid line is 
\begin{equation}
\kappa(\omega, T)^2=\kappa_0^2+(k_B T/c)^2 [1+(\hbar\omega/0.73 k_B T)^2],
\end{equation}
where $\kappa_0=0.044(1) \AA^{-1}$ and $c=4(1)\times 10^2$~meV$\AA$. At the QCP, $\kappa_0$ is expected to be zero, and its
value for La$_{1.95}$Sr$_{0.05}$CuO$_4$ is 
comparable to $\kappa_0$ in the superconducting state, but narrower than $\kappa_0$ at 40 K 
in the normal state for La$_{1.84}$Sr$_{0.16}$CuO$_4$\cite{lsco_nbc}. Hence, La$_{1.95}$Sr$_{0.05}$CuO$_4$ and the optimally doped LSCO with a short $1/\kappa_0$, about 6 Cu-Cu spacings, 
seem equally distant from the QCP. 

In conclusion, the Fincher-Burke modes, the broad and dispersive spin excitations of itinerant antiferromagnets, are observed in the spin-glass regime of La$_{2-x}$Sr$_x$CuO$_4$.
Theoretical understanding of similar excitation modes in superconducting cuprates now has an added task in the insulators. Befitting to the generality of its theoretical argument, the $\omega/T$ scaling is shown to be valid for a new type of cuprates above $T_X$. Spin excitations below $T_X$ remain gapless contrary to the prediction of non-random quantum theory. The spin-glass transition at finite doping introduces an extra component of slow spin fluctuations. It would be interesting to explore whether the glassy state\cite{bao05a,stock,gqcp}, limiting the correlation length of the rest of spins, is responsible for the equal distance from the QCP for La$_{1.95}$Sr$_{0.05}$CuO$_4$ and La$_{1.84}$Sr$_{0.16}$CuO$_4$.

Work at LANL is supported by U.S. DOE, and SPINS partially by NSF under Agreement No. DMR-0454672.


\end{document}